\pdfoutput=1
\RequirePackage{amssymb}
\RequirePackage[l2tabu,orthodox,abort]{nag}
\PassOptionsToPackage{final,stretch=10,spacing=true}{microtype}
\PassOptionsToPackage{hyphens,obeyspaces}{url}
\documentclass[sigconf]{acmart}
\usepackage[utf8]{inputenc}

\usepackage{enumitem}
\usepackage{fancyvrb}
\usepackage{graphicx}
\usepackage{listings}
\usepackage[detect-all]{siunitx}

\usepackage[repro]{reidpr}
\pdfid{8abb328d-367a-452c-87f8-19651313bd78}

\graphicspath{{images/}}
\setcitestyle{nocompress}  

\newcommand{\graphicsspace}{\vspace{-4pt}}
\newcommand{\listingspace}{\vspace{-6pt}}
\newcommand{\tablespace}{\listingspace}

\title{Minimizing Privilege for Building HPC Containers}

\author{Reid Priedhorsky}
\email{reidpr@lanl.gov}
\affiliation{%
  \department{High Performance Computing Division}
  \institution{Los Alamos National Laboratory}
  \city{Los Alamos}
  \state{NM}
  \country{USA}
}

\author{R.\ Shane Canon}
\email{scanon@lbl.gov}
\affiliation{%
  \department{National Energy Research Scientific Computing Center}
  \institution{Lawrence Berkeley National Laboratory}
  \city{Berkeley}
  \state{CA}
  \country{USA}
}

\author{Timothy Randles}
\email{trandles@lanl.gov}
\affiliation{%
  \department{High Performance Computing Division}
  \institution{Los Alamos National Laboratory}
  \city{Los Alamos}
  \state{NM}
  \country{USA}
}

\author{Andrew J.\ Younge}
\email{ajyoung@sandia.gov}
\affiliation{%
  \department{Center for Computing Research}
  \institution{Sandia National Laboratories}
  \city{Albuquerque}
  \state{NM}
  \country{USA}
}

\copyrightyear{Please cite the version of record: \url{https://doi.org/10.1145/3458817.3476187}}
\acmYear{}
\setcopyright{none}
\acmConference{Pre-print}{v3}{\today}
\acmBooktitle{}
\acmDOI{}
\acmISBN{}
\begin{document}

\begin{abstract}  


  HPC centers face increasing demand for software flexibility, and there is growing consensus that Linux containers are a promising solution. However, existing container build solutions require root privileges and cannot be used directly on HPC resources. This limitation is compounded as supercomputer diversity expands and HPC architectures become more dissimilar from commodity computing resources.
  %
  Our analysis suggests this problem can best be solved with low-privilege containers. We detail relevant Linux kernel features, propose a new taxonomy of container privilege, and compare two open-source implementations: mostly-unprivileged rootless Podman and fully-unprivileged Charliecloud.
  %
  We demonstrate that low-privilege container build on HPC resources works now and will continue to improve, giving normal users a better workflow to securely and correctly build containers.
  %
  Minimizing privilege in this way can improve HPC user and developer productivity as well as reduce support workload for exascale applications.

\end{abstract}

\maketitle

\section{Introduction}
\label{sec:introduction}

Linux containers have become a popular approach to develop, test, and deploy applications, because they let an application be packaged with its complete supporting software stack as a single unit, even if that stack is an OS distribution (the only exception being the kernel itself)~\cite{kamp2000jails, bernstein2014containers}. The simple foundation is that containers are processes (or groups of cooperating processes) with their own independent view of certain kernel resources, most importantly the filesystem tree.

Recent challenges for large-scale scientific applications in HPC include the need for greater flexibility in supporting software stacks as well as application teams' desire to control the timing of changes to their stacks. Containers are an increasingly popular way to meet these needs,\footnote{Containers for HPC system services is out of scope for this work.} and these benefits are not offset by a performance penalty. Studies of container performance impact have consistently shown that, when done correctly, containers introduce little to no performance penalty and sometimes even improve aspects like startup times~\cite{abraham2020containers, felter2015performance, liu2020performance, pedretti2020chronicles, torrez2019performance, sparks2017hpc, xavier2013performance}.

While containers have been deployed across a wide array of supercomputers in recent years~\cite{benedicic2019sarus, gerhardt2017shifter, le2017performance, torrez2019performance, younge2017tale}, containerized workloads still represent a distinct minority of supercomputer cycles~\cite{austin2020workload}. We believe a key reason is the difficulty of building container images with unprivileged tools.


A long-standing principle for HPC centers across all security levels
is to give users normal, unprivileged accounts, due to HPC's need for shared filesystems and their UNIX permissions. This “least-privilege” rule protects data from unauthorized access and the system from unauthorized modification, even if by mistake. In contrast, computing for industry web applications (a.k.a.\ “the cloud”) uses single-user virtual machine sandboxes without shared filesystems~\cite{qian2009cloud, wang2010cloud}. Data are protected by identity management, encryption, formal methods, and/or role-based authentication~\cite{cook2018formal, halabi2017quantification, hendre2015semantic}. Such isolation techniques make it secure to allow container runtimes and build tools with elevated privileges. A conflict arises when these tools are proposed for HPC resources, where elevated privilege is almost universally prohibited.

This paper explores container build privilege in HPC, discussing how containers work, different container implementations' privilege models and how they relate to Linux namespaces, and why \emph{running} unprivileged containers is easier than \emph{building} them. Next, we analyze potential approaches to enable container build on HPC and compare related work in this area. We then detail two container build solutions that have real, working implementations: rootless Podman's isolation of privileged operations into helper tools~\cite{walsh2021setup} and Charliecloud's fully unprivileged model~\cite{priedhorsky2017sc}.\footnote{Disclosure: Authors R.P.\ and T.R.\ are on the Charliecloud team.} We close with an analysis of how minimizing container build privileges can be further improved, as well as implications this new build capability has for future HPC workflows.

\section{Background}
\label{sec:background}


There are two common workflows in use today for container-based HPC application development. The first is for developers to build images on their local laptop or workstation, test on the same hardware, and then transfer the image to an HPC system for production runs. This model allows access to their preferred, local working environment, including editors, tools, configuration, etc. The second is a continuous integration / continuous deployment (CI/CD) approach, where images are automatically built on standalone and isolated resources, such as ephemeral virtual machines, upon code being pushed to a repository and passing tests. In both build workflows, privileged build is a reasonable choice.

However, these workflows raise problems for HPC. First, HPC systems have well-specified architectures that are increasingly diverse, whether a specific x86-64 microarchitecture (e.g., Haswell), a non-x86 CPU family (e.g., ARM or PowerPC), and/or accelerators of some kind (e.g., GPUs). HPC applications are usually performance-sensitive and therefore compiled for the specific architecture of the target supercomputer. On the other hand, developer workstations and CI/CD clouds are not well-specified and must be treated as generic x86-64 resources. This makes it difficult to build HPC applications properly on such resources. While in principle one could cross-compile, in practice this isn't adequately supported by HPC application build systems. Further, testing HPC applications is usually informative only on the target system.

A second problem is resources available only on specific networks or systems. Developers often need licenses for compilers, libraries, or other proprietary code with this limitation. Security-sensitive applications and data often have stringent restrictions on where they may be stored. This means container images must be built and stored in the same places.

In these and other scenarios, building on HPC resources closely matching the target architecture, perhaps supercomputers themselves, properly isolated and with direct access to services such as license servers, is the most natural approach. It can also accelerate development by eliminating the need to transfer images between networks. However, providing container image build capabilities on HPC resources is challenging at present.

The rest of this section explains why. Doing so first requires a deep dive into how containers work, specifically user namespaces and the correspondence between host user IDs (UIDs) or group IDs (GIDs) and those in the container. Based on this knowledge, we introduce a new taxonomy of privilege models for containers. Finally, we detail why naïve unprivileged container build doesn't work, to motivate the more complex approaches later in the paper.

\subsection{User namespaces and their implications}
\label{sec:userns}

Containers are currently implemented using Linux kernel features,\footnote{Container implementations on other operating systems work by adding Linux to that operating system, e.g. Docker on Mac's hypervisor approach~\cite{raina2018under}.} most importantly \vocab{namespaces}~\cite{kerrisk2013overview}, which let a process have its own view of certain kernel resources. Crucially, the \vocab{mount namespace} gives a process its own mounts and filesystem tree~\cite{man2020mountns}, allowing the container to run a different distribution than the host, and the \vocab{user namespace} gives a process its own UID and GID space~\cite{kerrisk2013userns, man2020userns}.

We give a focused discussion of namespaces, omitting three details: (1)~There are about a half dozen other types of namespace. (2)~If namespaces are compiled into the kernel, \emph{all} processes are within a namespace of every configured type whether or not containerized, even if some types are disabled via sysctl, which governs creation of \emph{new} namespaces. (3)~Some namespaces, including user, can be nested many levels deep. Our discussion uses a simple two-level host/container division, and we consider only the user and mount namespaces.

Though incomplete, this focused view is accurate and yields a more accessible discussion. Also, we mostly cover user namespaces because they are the key to low-privilege container implementations and have quite subtle behavior.


\subsubsection{UID and GID maps}
\label{sec:priv-userns}

The defining feature of user namespaces is that a process can have UIDs and GIDs inside a namespace that are different from its IDs outside. To accomplish this, the namespace is created with two one-to-one mappings: one between host UIDs and namespace UIDs, and another between host GIDs and namespace GIDs.\footnote{The kernel is concerned only with IDs, which are integers in the range $0$ to $2^{32}-1$ inclusive. Translation to usernames and group names is a user-space operation and may differ between host and container even for the same ID.} Because the map is one-to-one, there is no squashing of multiple IDs to a single ID in either direction. In both cases, it is the host IDs that are used for access control; the namespace IDs are just aliases. For example, one could map the unprivileged invoking host user to namespace UID~0 (i.e., root).\footnote{In-namespace UID~0 is special because the \code{execve(2)} system call that transfers control from container runtime to containerized application typically gives the process all capabilities within the namespace~\cite{man2020capabilities,man2020userns}. Thus, “UID~0” and “having all capabilities” can be treated the same for our discussion.} Benefits of this include (1)~giving a user root inside a namespace while remaining unprivileged on the host and (2)~allocating a set of IDs with well-known numbers (e.g., a distribution's system users and groups) that do not clash with the host. This containerized process would appear to be privileged within the namespace, but in reality is is just another unprivileged process. These mappings need not be complete. The four possibilities for a given pair of IDs are:

\begin{enumerate}

  \item \emph{In use on the host, and mapped to a namespace ID.} \label{map:inuse-mapped}
    The namespace ID is simply an alias of the host ID for use within the namespace.

  \item \emph{Not in use on the host, and mapped to a namespace ID.} This is the same as Case~\ref{map:inuse-mapped}, except the host ID happens to not yet be used. The kernel has no notion of “active” UIDs or GIDs, i.e., there is no initialization step for IDs. In particular, files can be owned by these IDs, but outside the namespace they will have no corresponding user or group names.

  \item \emph{In use on the host, but not mapped to a namespace ID.} These are valid inside the namespace, but there is no way to refer to them. Further, a user with access to a file via an unmapped group can still access it inside the namespace, even though \code{ls} will show it as \code{nogroup}. Confusingly, a file owned by a different unmapped group the user \emph{doesn't} have access to will \emph{also} be listed as \code{nogroup}. System calls and setuid/setgid executables do not accept unmapped IDs.

  \item \emph{Not in use on the host, and not mapped to a namespace ID.} These IDs are not available inside the namespace. Because they are not in use on the host, processes cannot have them when entering the namespace, and they can't be changed to within the namespace.

\end{enumerate}

%

While correct file ownership is critical in a multi-user system to protect users and processes from one another, multiple users and groups \emph{within an image} is rarely needed for HPC application containers. That is, if an HPC application is containerized, then anyone using the container should be able to see any and all files within it, because that is what's required to run the application. The concept of privileged and unprivileged users is not needed here. Making files accessible for only specific users or groups isn't useful in this context, and could lead to inconsistent behavior depending on who is running the containerized application. This is also true for many containerized HPC services such as databases, which can run as the same user as the jobs they support.

To be clear, there are scenarios where containerized applications or application-support tools do need to run with multiple users or with different privilege levels, e.g. some web services or databases that need to act on behalf of multiple users. However, these are the exception rather than the norm for HPC, and require close sysadmin involvement whether containerized or not.

\subsubsection{Privileged ID maps}

Processes that hold \code{CAP_SETUID} or \code{CAP_SETGID}~\cite{man2020capabilities} (and UID~0 typically holds all capabilities) can set up mostly-arbitrary UID and GID maps respectively. This privileged functionality\footnote{While \code{CAP_SETUID} is limited privilege, such a process can gain additional privileges by changing its UID to zero, so it's critical to ensure no such execution paths exist.} can be used in otherwise-unprivileged implementations via setuid or setcap helper programs to set up the user namespace. Currently, the standard helpers are \code{newuidmap(1)} and \code{newgidmap(1)} from the shadow-utils package~\cite{man2020newuidmap, man2020newgidmap}, though some implementations do provide their own (e.g., Singularity).

\begin{figure}
  \includegraphics[width=\columnwidth]{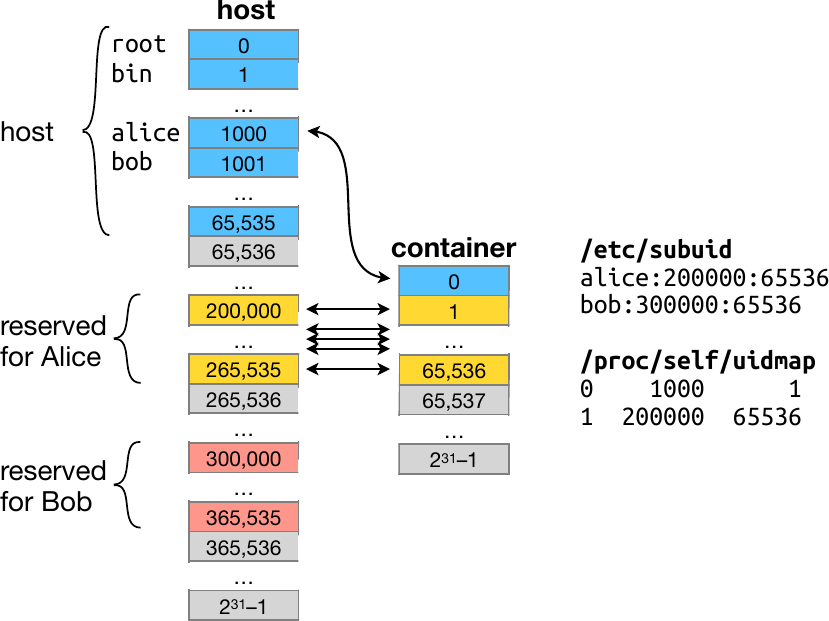}%
  \graphicsspace
  \caption{Typical privileged UID map for container run by Alice. The file \protect\code{/etc/subuid} configures the user-space helper for host UIDs Alice and Bob may use; \protect\code{/proc/self/uidmap} is the subsequent kernel mapping.}
  \label{fig:uidmap}
\end{figure}

A typical configuration for image building is to map the calling user to root, and then map a sequence of unused host UIDs to container UIDs 1–$n$; \figurename~\ref{fig:uidmap} shows an example. GIDs have an analogous map. This yields valid in-namespace IDs from 0 to $n$, with $n$ selected to cover the distribution's system users and groups. Linux's 32-bit IDs help provide enough IDs to everyone who needs to build container images. However, an important task for sysadmins is to ensure that all host IDs really are unused; for example, it's easy to miss IDs on unmounted NFS shares.


The privileged helper approach does need to be implemented carefully, for two reasons. First, the helper is a security boundary, responsible for ensuring that unprivileged users can set up only safe maps; we are aware of real-world errors in such checks, making it a practical as well as theoretical concern. Thus, standard helpers provided by the host OS distribution are preferred, as more rigorous security review and patching is available. Second, sysadmins must correctly configure permissible ID mappings. For example, in \figurename~\ref{fig:uidmap}, if host UID~1001 mapped to container UID~65,537, Alice would have access to all of Bob's files. She could use her container root to \code{setuid(65537)}, which maps back to host UID~1001, and there are many ways to access the host's filesystem tree from a container. We emphasize the importance of correct configuration here. System management is already complicated, and when new, unfamiliar features are added, the risk of human error increases.

An important corollary of the ID map is that IDs will be correct only within the container. Even though image filesystem trees can be easily available both inside and outside the container, file IDs on the outside will be the mostly-arbitrary host side of the map. For example, images are often stored in  tar archives. With privileged ID maps, for correct IDs these archives must be created within the container or use an ID source other than the filesystem.

\subsubsection{Unprivileged ID maps}

Unprivileged processes can make only limited mappings: (1)~host UID to arbitrary in-namespace UID and (2)~host GID to arbitrary in-namespace GID~\cite{man2020userns}. Supplementary groups must remain unmapped. Thus, the process has precisely the same access within the container as on the host.

Because they are unmapped, supplementary groups have reduced functionality:  \code{chgrp(1)} can't be used on them, and they can cause confusion when displayed as \code{nogroup}.

In other words, for unprivileged user namespaces, the ID map does not matter, except for programs that check for specific IDs. Otherwise, mapped IDs in the unprivileged case are simply a convenience provided by the kernel, with only cosmetic effects.

\subsubsection{The \protect\code{setgroups(2)} trap}

This system call requires careful handling when setting up a privileged user namespace due to obscure interactions between the GID map and UNIX permissions. It lets a process add or remove supplementary groups~\cite{man2019getgroups}, which have different implications in a user namespace.

Adding groups to a process is normally uninteresting from an access control perspective, because it is a privileged system call, i.e. the process already has access to everything. In privileged user namespaces, this operation is more consequential because they can choose whether to enable \code{setgroups(2)}. If enabled, anyone with root in the namespace also has access to everything protected by all mapped groups. The lesson is the same as the prior section: sysadmins must configure allowed group mappings carefully.

\emph{Removing} groups from a process has an interesting subtlety. UNIX permissions are evaluated in the order user, group, other — and the first match governs~\cite{man2020userns}. For example, consider a production system containing a file \code{/bin/reboot} owned by \code{root:managers} and permissions \code{rwx---r-x}. With this setup, all users \emph{except managers} can reboot the system. However, any manager process able to call \code{setgroups(2)} can drop \code{managers} from its groups, changing the match from group to other, and now that manager can reboot the system. Without user namespaces, this situation again cannot arise, because well-run sites do not give managers root. Privileged helpers are responsible for disabling \code{setgroups(2)} when acting on behalf of unprivileged users; \code{newgidmap(1)} did have a vulnerability when it failed to do so~\cite{cve-2018-7169}.

Neither of these issues are concerning for unprivileged user namespaces because \code{setgroups(2)} is not available~\cite{man2020userns}.

\subsection{Container privilege levels}

This paper uses the term \vocab{privileged} for processes that run with escalated privilege, whether root or non-trivial capabilities~\cite{man2020capabilities}. The terms \vocab{unprivileged} or \vocab{fully unprivileged} refer to processes that run as normal users with no capabilities and \vocab{mostly unprivileged} for such processes that call privileged (setuid or setcap) helper programs; \vocab{low privilege} refers to either fully or mostly unprivileged.

We propose a three-level taxonomy of container privilege, based on the presence or absence of the user namespace and whether it is created with host privileges:

\begin{enumerate}[label=Type~\Roman*., ref=\Roman*]

\item \label{ct:priv-nouser}
  Mount namespace (or \code{chroot(2)}) but no user namespace. Privileged setup; shares IDs with the host system. Root inside a container is root on the host.

\item \label{ct:priv-user}
  Mount namespace and privileged user namespace. Privileged setup; arbitrarily many UIDs and GIDs independent from the host. Root inside the container is typically mapped to an unprivileged host user, not root.

\item \label{ct:unpriv-user}
  Mount namespace and unprivileged user namespace. Unprivileged setup; only one UID and one GID mapped into the container, which are aliases of the invoking user's IDs. That is, containerized processes remain unprivileged, holding the user's normal unprivileged IDs.

\end{enumerate}

A well-known example of a Type~\ref{ct:priv-nouser} implementation is Docker~\cite{avram2013docker}. To demonstrate Type~\ref{ct:priv-user} and~\ref{ct:unpriv-user}, this paper uses rootless Podman~\cite{henry2019podman, walsh2021setup} and Charliecloud~\cite{priedhorsky2017sc} respectively. We expect lessons derived will be applicable to other implementations as well.

\subsection{Unprivileged container build is hard}
\label{sec:why-hard}

Building a container image largely entails running a sequence of containerized commands on a read-write image. This currently works much like building a system: unpack a Linux distribution base, install the necessary distribution packages (e.g., with \code{dpkg} or \code{rpm}), and then install the desired application(s). The difference from \emph{running} containers is that distribution package managers assume privileged access, and key packages need multiple UIDs/GIDs and privileged system calls like \code{chown(2)} to install. We have not yet encountered any fundamental reason privileged access is needed; the tools install software, rather than using privileged operations like listening on low ports. However, multiple decades of Linux software delivery has done it this way, and it works well. There has not yet been a need to change.

\begin{figure}
\begin{lstlisting}
$ cat centos7.dockerfile
FROM centos:7
RUN echo hello
RUN yum install -y openssh
$ ch-image build -t foo -f centos7.dockerfile .
  1 FROM centos:7
  2 RUN ['/bin/sh', '-c', 'echo hello']
hello
  3 RUN ['/bin/sh', '-c', 'yum install -y openssh']
[...]
  Installing : openssh-7.4p1-21.el7.x86_64
Error unpacking rpm package openssh-7.4p1-21.el7.x86_64
error: unpacking of archive failed on file [...]: cpio: chown
[...]
error: build failed: RUN command exited with 1
\end{lstlisting}
  \listingspace
  \caption{Simple CentOS 7-based Dockerfile that fails to build in a basic Type~\ref{ct:unpriv-user} container because \protect\code{chown(2)} failed. We selected the OpenSSH client because it's problematic across distributions and common in HPC user containers. (This and later transcripts have been simplified for clarity. The notation \protect\code{[...]} is sometimes used to avoid confusion, but in most cases irrelevant text is simply omitted.)}
  \label{fig:centos-fail}
\end{figure}

Ideally, we would be be able to install packages using a fully unprivileged (Type~\ref{ct:unpriv-user}) container implementation with no special modifications or alterations. Unfortunately, this is not the case. For example, \figurename~\ref{fig:centos-fail} presents a simple CentOS-based Dockerfile that fails to build in such a configuration. In this case, the system call \code{chown(2)} fails because while the process appears to be root within the container, it is really an unprivileged process running as the invoking user, and thus may not call \code{chown(2)}~\cite{man2020chown}.

\begin{figure}
\begin{lstlisting}
$ cat debian10.dockerfile
FROM debian:buster
RUN echo hello
RUN apt-get update
RUN apt-get install -y openssh-client
$ ch-image build -t foo -f debian10.dockerfile .
  1 FROM debian:buster
  2 RUN ['/bin/sh', '-c', 'echo hello']
hello
  3 RUN ['/bin/sh', '-c', 'apt-get update']
E: setgroups 65534 failed - setgroups (1: Operation not permitted)
E: setegid 65534 failed - setegid (22: Invalid argument)
E: seteuid 100 failed - seteuid (22: Invalid argument)
[...]
error: build failed: RUN command exited with 100
\end{lstlisting}
  \listingspace
  \caption{Simple Debian 10-based Dockerfile that fails to build in a basic Type~\ref{ct:unpriv-user} container. Here, \protect\code{apt-get} encountered errors trying (ironically) to drop privileges.}
  \label{fig:debian-fail}
\end{figure}

This problem is not unique to RPM-based distributions. In \figurename~\ref{fig:debian-fail}, a Debian-based Dockerfile also fails to build. In this case, \code{apt-get} tries to drop privileges and change to user \code{_apt} (UID~100) to sandbox downloading and external dependency solving~\cite{dicarlo2019release, wienemann2021apt}. This yields a series of failed system calls. \code{setgroups(2)} fails because that system call is not permitted in an unprivileged container~\cite{man2020userns}.  \code{setresgid(2)} (not \code{setegid(2)} and \code{seteuid(2)} as in the error messages) fails twice because again \code{apt-get} is actually unprivileged and may not make those system calls~\cite{man2017setresuid}.


The colliding assumptions of (1)~package managers require root and (2)~unprivileged containers disallow root mean that unprivileged container build is a harder problem than unprivileged run. The next section outlines solutions.

\section{Approaches for container build}

Above, we've motivated the need for unprivileged container build in HPC and detailed how user namespaces can provide reduced-privileged containers. In this section, we take a step back, assessing existing container implementations used in HPC and approaches for limiting build privilege. This assessment reinforces our conclusion that user namespace-based container builds are currently the lowest-cost path forward for HPC.

\subsection{Container implementations used in HPC}
\label{sec:related}

There are several container implementations in active or potential use for HPC. Containers originated in the web applications world~\cite{pahl2016microservices}, which often uses virtual machine sandboxes as noted above, so running applications in a Type~\ref{ct:priv-nouser} container is a reasonable choice whether or not the containerized processes are privileged. Also, user namespaces were not available until Linux~3.8, released in February~2013~\cite{torvalds2013linux38}, and they were not fully supported in Red Hat Enterprise Linux (RHEL) or its derivatives, upon which most HPC centers rely, until version 7.6 in October 2018~\cite{redhat2018release}. Without user namespaces, only Type~\ref{ct:priv-nouser} containers are possible.

Docker was the first container implementation to become widely popular. Its initial public release was March~2013 and supported Linux~2.6.24~\cite{avram2013docker}, making it Type~\ref{ct:priv-nouser} by necessity. Even simply having access to the “\code{docker}” command is equivalent to root “by design”~\cite{reventlov2015root}. Another key design decision was its client-daemon execution model~\cite{henry2019podman}. Processes started with “\code{docker run}” are descendants of the Docker daemon, not the shell, which is undesirable for HPC because it is another service to manage/monitor, breaks process tracking by resource managers, and can introduce performance jitter~\cite{ferreira2008characterizing}. Docker quickly became the industry standard, and in fact many containers intended for supercomputers are currently built with Docker using the workflows noted above. Because its approach worked fine for web applications, it gained inertia and mind share even as unprivileged kernel container features matured. Docker did add a Type~\ref{ct:priv-user} mode in 2019~\cite{docker2021rootless}, though it is not the default and to our knowledge is not yet widely used in practice.


The HPC community has also produced container implementations, some with build capability. The most popular HPC container implementation is currently Singularity~\cite{kurtzer2017singularity}, which can run as either Type~\ref{ct:priv-nouser} or \ref{ct:priv-user} (branded “fakeroot”). As of this writing, Singularity~3.7 can build in Type~\ref{ct:priv-user} mode, but only from Singularity definition files. Building from standard Dockerfiles requires a separate builder (e.g., Docker) followed by conversion to Singularity's image format, which is a limiting factor for interoperability. Type~\ref{ct:priv-nouser} HPC examples include both Shifter~\cite{jacobsen2015cug} and Sarus~\cite{benedicic2019sarus}, though currently these focus on distributed container launch rather than build. Another implementation of interest is Enroot, which advertises itself as “fully unprivileged”~\cite{abecassis2020containers} with “no setuid binary”~\cite{calmels2021enroot}, i.e., Type~\ref{ct:unpriv-user}. However, as of the current version 3.3, it does not have a build capability, relying on conversion of existing images.


%
%

\subsection{Summary of build options}

In this section, we propose some requirements for container build and analyze three general approaches in light of them. First, the build recipe (typically a Dockerfile) should require no modifications. Because existing recipes are often used in whole or in part, modifications impact productivity and portability, which are major benefits of containers. Second, we treat the current behavior of distribution packaging tools as unchangeable. (We discuss below in §\ref{sec:rec-distros} how these tools' privilege needs could be relaxed.) With these requirements in mind, we consider three approaches:

\begin{enumerate}

\item \emph{Sandboxed build system}. \label{it:sandbox}
  One way to work around increased privileges is to create an isolated environment specifically for image builds. Many approaches are available, including most commonly virtual machines or bare-metal systems with no shared resources such as production filesystems. When creating these environments, the risk of users circumventing controls must be carefully minimized. These models do work well with CI and other automated approaches, where further limits may be available.

\item \emph{Type~\ref{ct:priv-user} containers}. \label{it:priv-user}
  With the isolated UIDs and GIDs provided by user namespaces, this approach can provide a mostly-seamless illusion of root within the container while remaining mostly unprivileged. Privileged helper tools are needed to set up the user namespace, and sites must maintain correct mapping files to ensure users are properly isolated from each other.

\item \emph{Type ~\ref{ct:unpriv-user} containers.} \label{it:unpriv-user}
  This type of container is fully unprivileged, but limited UIDs and GIDs mean more is needed to install arbitrary packages. One solution is a wrapper like \code{fakeroot(1)}, which intercepts privileged system calls and fakes their success. This wrapper can be injected automatically to meet the first requirement above.

\end{enumerate}

Recall that building containers with low privilege on HPC resources is the focus of this work. Option~\ref{it:sandbox} offers a pragmatic way to work around privilege issues and is already in use at many sites, e.g., via the Sylabs Enterprise Remote Builder~\cite{sylabsremote}. However, isolated build environment may not be able to access needed resources, such as private code or licenses. Thus, our view is that Options~\ref{it:priv-user} and~\ref{it:unpriv-user} are the most promising.

These two options are distinguished by user namespaces being privileged or unprivileged, respectively. Aside from details of any specific implementation, there are two key factors to assess this distinction. First, are the security challenges of privileged namespace setup helpers and their configuration acceptable to a site? Second, how much does a site care that container images include specific or multiple IDs? If the helpers are acceptable and specific/multiple IDs are needed for exact container images, then sites should consider Option~\ref{it:priv-user}, detailed below in~§\ref{sec:podman} using rootless Podman. Otherwise, they should consider Option~\ref{it:unpriv-user}, detailed in ~§\ref{sec:charliecloud} using Charliecloud.

%

\section{Podman}
\label{sec:podman}

As detailed above, one way to provide a reasonable container build capability for HPC resources is by using a Type~\ref{ct:priv-user} container build solution directly on a supercomputer. A recent container implementation under this cateogory is Podman,\footnote{\url{https://podman.io}} and in particular the “rootless Podman” configuration, which is particularly relevant to HPC \cite{walsh2018reintroduction, walsh2021setup}. 
The main design goals of rootless Podman are to have the same command-line interface (CLI) as Docker, remove root-level privileges, and use a fork-exec model rather than Docker's client-daemon model~\cite{henry2019podman}. Because Podman is CLI equivalent to Docker, many users can successfully utilize Podman by \code{alias docker=podman} and use as expected. Podman also adheres to the OCI spec for container compatibility and interoperability, making transition to Podman from existing container implementations effectively seamless. While Podman is open source, it is heavily supported by Red Hat and integrated into their Linux distribution as of RHEL7.8.

While we refer to rootless Podman, or Podman more generally as a Type~\ref{ct:priv-user} container implementation in this paper, the implementation is in fact based on the same code as the Buildah tool and other related container utilities. Podman in this sense only provides a CLI interface identical to Docker, whereas Buildah provides more advanced and custom container build features. As Podman and Buildah leverage the same codebase for build operations, we can assume they are functionally equivalent in this paper for purposes of brevity. 
%
As Podman avoids Docker's daemon design, it has gained significant interest for HPC applications as it is a much better fit for HPC deployments where daemons are generally discouraged whenever possible. While this design decision can seem inconsequential, the removal of a daemon can have significant improvements to overall system performance and manageability \cite{petrini2003case}.

\subsection{Rootless Podman with privileged helpers}
\label{sec:rootlesspodman}

As detailed in Section \ref{sec:userns}, the ability to minimize privilege with rootless Podman is a key feature for HPC. In actuality, rootless Podman refers to the ability to build and run containers without administrative privileges. While there are other potential configurations, the most common implementation currently uses Privileged ID maps to help mitigate the security issues of Docker's inherent privilege escalation. Effectively, this is a Type ~\ref{ct:priv-user} configuration.  Podman itself remains completely unprivileged; instead a set of carefully managed tools provided by shadow-utils are executed by Podman to set up user namespaces mappings. The shadow-utils executables, \code{newuidmap} and \code{setgidmap}, map the specified UIDs and GIDs into the container's user namespace. These tools read \code{/etc/subuid} and \code{/etc/subgid}, respectively, which outline the user namespace mappings.  The shadow-utils executables are installed using \code{CAP_SETUID}, which helps minimize risk of privilege escalation compared to using a \code{SETUID} bit.  While this Linux capability still poses a potential security risk, the multiple points of privilege separation between Podman and shadow-utils, along with the careful validation of the executables' implementation by a trusted Linux OS distribution (such as Red Hat), can reasonably minimize the possibility of escalation to the point where we beleive Podman can be safely used in a production environment. Furthermore, Podman can be additionally secured with SELinux to further prevent potential points of exploitation.   
%

\begin{figure}
\begin{lstlisting}
$ cat /etc/subuid
# USER : STARTUID : TOTALUIDs
alice:200000:65536
bob:300000:65536
$ podman unshare cat /proc/self/uid_map
         0      1234          1
         1     200000      65536
\end{lstlisting}
  \listingspace
  \caption{An example subuid file outlining the user namespace mappings used by Podman, followed by a listing of the UID map used by Podman. The unshare command illustrates how user namespace mapping is mapping UID 0 in the container to 1234 on the host for a range of 1 UIDs. While not listed, subgid file would have similar mappings.}
  \label{fig:subuid}
\end{figure}

An example namespace mapping is provided in \figurename~\ref{fig:subuid}. Here, you can see the user \emph{alice} is able to allocate 65535 UIDs, starting at UID 200000. The user namespace mapping definitions cannot exceed the maximum user namespaces as configured in \code{/proc/sysuser/max_user_namespaces}. 
While the approach of privileged helpers works well in practice, it needs to be implemented carefully as the setuid helper is responsible for enforcing a security boundary to ensure unprivileged users can set up only safe maps.  These mappings need to be  specified by the administrator upon Podman installation for existing users.  Newer versions of shadow-utils can automatically manage the setup using \code{useradd} and \code{usermod --add-subuids}, which further simplifies administrative burden and lowers the risk of conflicts.  This ability for system administrators to carefully control user access to Podman's capabilities can be an asset for facilities looking to first prototype rootless Podman without fully enabling the utility system-wide. 

Podman also provides several additional features to provide a full-featured and production container solution to users. 
First, Podman by default leverages the runc container runtime for OCI compliance. This usage of runc as the underlying base runtime is similar to the implementation of Sarus~\cite{benedicic2019sarus} and helps conform to OCI standards to provide interoperability and compatibility with not only other container implementations, but also many OCI-compliant container registry services. 
Second,
Podman uses the fuse-overlayfs storage driver which provides unprivileged mount operations using a fuse-backed overlay file-system. 
Podman can also use the VFS driver, however this implementation is much slower and has significant storage overhead, and should only be used when absolutely necessary with existing installations (eg: RHEL7).

With rootless Podman, cgroups are left unused as cgroup operations by default are generally root-level actions. This is a convenient coincidence for HPC, since the on-node resource management is typically handled by the resource manager and job scheduler and having multiple services trying to interact with cgroups should be avoided. There still may be future utility in cgroups operations, so prototype work is underway to implement cgroups v2 in userspace via the crun runtime, which enables cgroups control in a completely unprivileged context. 
In summary, rootless Podman with shadow-utils privileged mappers to help manage user namespace mappings, provide a full-featured and Docker CLI-equivalent container build solution on HPC resources. When Podman is configured appropriately, the examples detailed in \figurename{}s~\ref{fig:centos-fail} and \ref{fig:debian-fail} will both succeed as expected when executed by a normal, unprivileged user.

%
%

\subsubsection{Unprivileged Podman}

While the default rootless Podman usage is designed around the privileged mappers from shadow-utils to provide a comprehensive user namespace mapping, it is also possible to set up Podman in an experimental unprivileged mode. Instead of the namespace mappers as illustrated in \figurename~\ref{fig:subuid}, the subuid mappers are discarded and a single UID is mapped to the container. When the UID mappers for a given user are left unset and the \code{--ignore_chown_errors} option is enabled with either the VFS or overlay driver, Podman then operates with unprivileged namespaces with one UID mapping, as illustrated in Figure \ref{fig:podman1namespace}.
However, the \code{yum install openssh-server} example in Section \ref{sec:background} will fail because /proc and /sys mappings in the container are owned by user \code{nobody}; a consequence of the namespace mapping. 
While further workarounds with \code{fakeroot(1)} as detailed in the following section could be applied here, this setup is not yet used in practice.


\begin{figure}
\begin{lstlisting}
$ cat /etc/subuid
$ podman unshare cat /proc/self/uid_map
         0      1234          1
\end{lstlisting}
  \listingspace
  \caption{Podman UID mapping in unprivileged mode. This setup does not use privileged helpers but has limitations.}
  \label{fig:podman1namespace}
\end{figure}

\subsection{Podman on the Astra Supercomputer}

To enable user-driven container build capabilities directly on HPC resources, rootless Podman was first deployed and tested on the Astra supercomputer \cite{pedretti2020chronicles}. Astra has proven to be an ideal first system to prototype the use of Podman for two reasons. First and most critically, Astra was the first Arm-based supercomputer on the Top 500 list, which meant existing containers based on \code{x86_64} would not execute on Astra. Instead, users had an immediate need to \emph{build} new container images specifically for the \code{aarch64} ISA, and in particular the Marvell Thunder X2 CPUs. This requirement included a containerizing ATSE, the Advanced Tri-Lab Software Environment, which provides an open, modular, extensible, community-engaged, and vendor-adaptable software ecosystem for which many production HPC codes from NNSA labs depend on~\cite{younge2020atse}. Second, Astra was initially deployed as a prototype system, which gave researchers additional latitude to test and evaluate experimental system software capabilities like those found with Podman. 

\begin{figure}
  \includegraphics[width=\columnwidth]{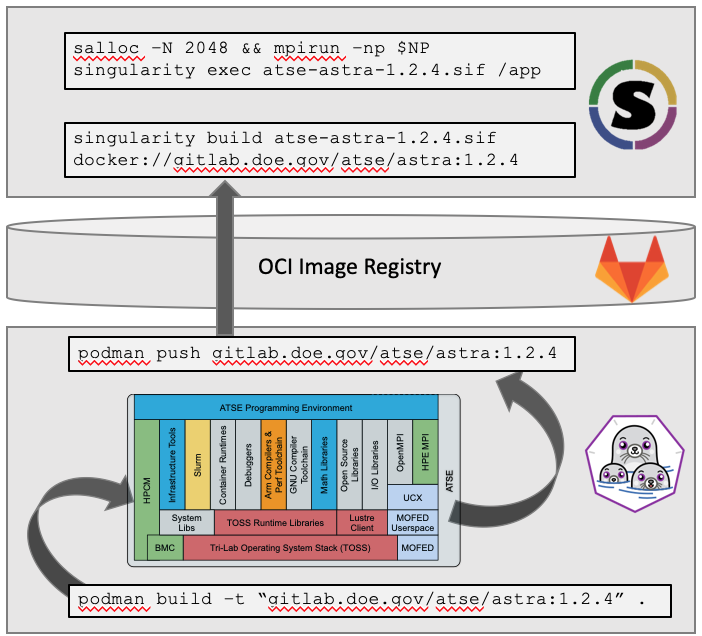}%
  \graphicsspace
  \caption{Container Build Workflow on Astra, with Podman}
  \label{fig:astra-containers}
\end{figure}

With Astra, Podman was used as part of a multi-stage container DevOps process and is outlined in \figurename~\ref{fig:astra-containers}. First, \code{podman build} is invoked from the Astra login node to construct the ATSE container build process in a directory which includes the Dockerfile for ATSE. This builds all of the ATSE components, including compilers, MPI libraries, third-party libraries, and test applications all directly in a container on the login node. This command is invoked by the user. Once the build is completed, it can be pushed to a OCI-compliant container registry (in this case the Gitlab Container Registry Service). A container registry is important to leverage in this workflow as it provides persistence to container images which could help in  portability, debugging with old versions, or general future reproducibility.   Finally, the container image built on the supercomputer can be deployed in parallel using the local resource management tool and an HPC container runtime. 
This was originally demonstrated with Singularity, however any HPC container runtime such as Charliecloud or Shifter could also be used with a similar step. While Podman could potentially be used for parallel execution \cite{gantikow2020rootless}, it has not yet been optimized for distributed container execution and likely would introduce significant overhead at scale.

To our knowledge, Astra's usage of Podman version 1.4.4 (Buildah 1.9.0) is the first demonstration of unprivileged users building OCI containers directly on a supercomputer. However, the implementation is not without limitations.  First, Podman with RHEL7.6 used the VFS driver, which is known to add overhead. 
Newer versions of Podman, as found in RHEL8, can leverage the overlayfs-fuse driver for better performance.
Second, the UID/GID mappers cannot work when the container storage location is a shared filesystem, such as NFS.  The reason is that the filesystem server has no way to enforce the file creation of different UIDs on the server side and cannot utilize the user namespace configuration. For Astra, either \code{/tmp} or local disk can be used for container storage on the login nodes, which requires additional configuration by users and does not enable any distributed launch abilities.

\section{Charliecloud}
\label{sec:charliecloud}

Charliecloud\footnote{\url{https://hpc.github.io/charliecloud}} is a container implementation designed for HPC applications with emphasis on the unprivileged use cases discussed in the first half of this paper~\cite{priedhorsky2017sc}. It is distinguished from other implementations because it is lightweight (roughly 4,000 lines of code as tested, version 0.23) and has been Type~\ref{ct:unpriv-user} from its first release. Its runtime \code{ch-run(1)} is written in C, and it has a Dockerfile interpreter and container registry API client written in Python, \code{ch-image(1)}. To our knowledge, this is the first fully unprivileged builder able to build Dockerfiles that make privileged system calls, which it does by automatically injecting \code{fakeroot(1)} into the build. This section details the approach. We first present background on \code{fakeroot(1)} and show how it can work around the limitations of unprivileged container build; we then describe Charliecloud's injection of the wrapper, which should be transferrable to any Type~\ref{ct:unpriv-user} implementation.

\begin{figure}
\begin{lstlisting}
$ fakeroot ./fakeroot.sh
+ touch test.file
+ chown nobody test.file
+ mknod test.dev c 1 1
+ ls -lh test.dev test.file
crw-r----- 1 root   root 1, 1 Feb 10 18:09 test.dev
-rw-r----- 1 nobody root    0 Feb 10 18:09 test.file
$ ls -lh test*
-rw-r----- 1 alice  alice  0 Feb 10 18:09 test.dev
-rw-r----- 1 alice  alice  0 Feb 10 18:09 test.file
\end{lstlisting}
  \listingspace
  \caption{Example of \protect\code{fakeroot(1)} use. This script changes ownership of a file and then creates a device file, both of which are privileged operations. \protect\code{fakeroot(1)} intercepts the system calls and causes them to “succeed”. Within the \protect\code{fakeroot(1)} context, \protect\code{ls(1)} shows the expected results (device file and \protect\code{nobody}-owned file, respectively); the subsequent unwrapped \protect\code{ls(1)} exposes the lies.}
  \label{fig:fakeroot}
\end{figure}

\subsection{Background: Faking root with \protect\code{fakeroot(1)}}

\code{fakeroot(1)} is a program to run a command in an environment that appears to be privileged but is not~\cite{man2019fakeroot}.\footnote{Not to be confused with Singularity's Type~\ref{ct:priv-user} mode, also called “fakeroot”.} It does this by intercepting privileged and privileged-adjacent system calls and lying to the wrapped process about their results. For example, an unprivileged process could \code{chown(2)}, and \code{fakeroot(1)} will return success from its own implementation without ever actually calling the system call. It also remembers which lies it told, to make later intercepted system calls return consistent results. In this example, fake results for \code{stat(2)} include the user and/or group set by the earlier fake \code{chown(2)}. Because \code{stat(2)} is not privileged, the wrapper really will make the system call, but then adjust the results to make them look correct in the “privileged” context. \figurename~\ref{fig:fakeroot} shows an example of \code{fakeroot(1)} use.

\code{fakeroot(1)} is not a perfect simulation but rather just enough to work for its intended purpose, which is building distribution packages, allowing “users to create archives (tar, ar, .deb etc.) with files in them with root permissions/ownership”~\cite{man2019fakeroot}. The focus is on filesystem metadata; e.g., one can't use \code{fakeroot(1)} to listen on an unprivileged port and make a process see a privileged port. Fortunately, this purpose is highly congruent with the needs of unprivileged container build.

\begin{table*}
  \centering
  \begin{tabular}{@{}lcccccc@{}}
    \toprule
      \textbf{implementation}
    & \textbf{initial release}
    & \textbf{latest version}
    & \textbf{approach}
    & \textbf{architectures}
    & \textbf{daemon?}
    & \textbf{persistency}
    \\
    \cmidrule(r){1-1}
    \cmidrule(l){2-7}
      fakeroot~\cite{fakeroot-gitlab,man2019fakeroot,man2019faked}
    & 1997-Jun
    & 2020-Oct (1.25.3)
    & \code{LD_PRELOAD}
    & any
    & yes
    & save/restore from file
    \\
      fakeroot-ng~\cite{fakeroot-ng-sourceforge,shachar2019fakeroot-ng-wiki}
    & 2008-Jan
    & 2013-Apr (0.18)
    & \code{ptrace(2)}
    & PPC, x86, x86-64
    & yes
    & save/restore from file
    \\
      pseudo~\cite{pseudo-git,man2020pseudo}
    & 2010-Mar
    & 2018-Jan (1.9.0)
    & \code{LD_PRELOAD}
    & any
    & yes
    & database
    \\
    \bottomrule \\
  \end{tabular}
  \tablespace
  \caption{Summary of \protect\code{fakeroot(1)} implementations.}
  \label{tab:fakeroots}
\end{table*}

There are three \code{fakeroot(1)} implementations we are aware of, summarized in Table~\ref{tab:fakeroots}. They each have different quirks; for example, \code{LD_PRELOAD} implementations are architecture-independent but cannot wrap statically linked executables, while \code{ptrace(2)} are the reverse. We've encountered packages that fakeroot cannot install but fakeroot-ng and pseudo can; fakeroot appears to be under more active development.


\subsection{Fully unprivileged build via \protect\code{fakeroot(1)}}

\begin{figure}
\begin{lstlisting}
$ cat centos7-fr.dockerfile
FROM centos:7
RUN yum install -y epel-release
RUN yum install -y fakeroot
RUN echo hello
RUN fakeroot yum install -y openssh
$ ch-image build -t foo -f centos7-fr.dockerfile .
  1 FROM centos:7
  2 RUN ['/bin/sh', '-c', 'yum install -y epel-release']
[...]
Complete!
  3 RUN ['/bin/sh', '-c', 'yum install -y fakeroot']
[...]
Complete!
  4 RUN ['/bin/sh', '-c', 'echo hello']
hello
  5 RUN ['/bin/sh', '-c', 'fakeroot yum install -y openssh']
[...]
Complete!
grown in 5 instructions: foo
\end{lstlisting}
  \listingspace
  \caption{CentOS 7 Dockerfile from \figurename~\ref{fig:centos-fail}, modified to build successfully by wrapping the offending \protect\code{yum install} with \protect\code{fakeroot(1)}.}
  \label{fig:centos-win-manual}
\end{figure}

We can use \code{fakeroot(1)} to fix the failed container builds in §\ref{sec:why-hard}. This approach will squash the actual ownership of all files installed to the invoking user, but that usually does not matter for HPC applications, and any downstream Type~\ref{ct:unpriv-user} users that pull the image will change ownership to themselves anyway, like \code{tar(1)}~\cite{info2021tar}.

\figurename~\ref{fig:centos-win-manual} shows modifications of the CentOS~7 Dockerfile from \figurename~\ref{fig:centos-fail} to build it under \code{ch-image}'s simplest configuration. Three changes were needed:
\begin{enumerate}

\item Install EPEL, because CentOS has no \code{fakeroot} package in the default repositories.

\item Install \code{fakeroot}.

\item Prepend \code{fakeroot(1)} to the \code{openssh} install command.

\end{enumerate}
The first two install steps do use \code{yum(1)}, but fortunately these invocations work without \code{fakeroot(1)}.

\begin{figure}
\begin{lstlisting}
$ cat debian10-fr.dockerfile
FROM debian:buster
RUN echo 'APT::Sandbox::User "root";' > /etc/apt/apt.conf.d/no-sandbox
RUN echo hello
RUN apt-get update
RUN apt-get install -y pseudo
RUN fakeroot apt-get install -y openssh-client
$ ch-image build -t foo -f debian10-fr.dockerfile .
  1 FROM debian:buster
  2 RUN ['/bin/sh', '-c', 'echo \'APT::Sandbox::User "root";\' >                           /etc/apt/apt.conf.d/no-sandbox']
  3 RUN ['/bin/sh', '-c', 'echo hello']
hello
  4 RUN ['/bin/sh', '-c', 'apt-get update']
[...]
Fetched 8422 kB in 7s (1214 kB/s)
Reading package lists...
  5 RUN ['/bin/sh', '-c', 'apt-get install -y pseudo']
[...]
Setting up pseudo (1.9.0+git20180920-1) ...
[...]
W: chown to root:adm of file /var/log/apt/term.log failed [...]
  6 RUN ['/bin/sh', '-c', 'fakeroot apt-get install -y                                     openssh-client']
[...]
Setting up openssh-client (1:7.9p1-10+deb10u2) ...
Setting up libxext6:amd64 (2:1.3.3-1+b2) ...
Setting up xauth (1:1.0.10-1) ...
Processing triggers for libc-bin (2.28-10) ...
grown in 6 instructions: foo
\end{lstlisting}
  \listingspace
  \caption{Debian 10 Dockerfile from \figurename~\ref{fig:debian-fail}, modified by configuring \protect\code{apt-get(1)} to not drop privileges and wrapping the offending \protect\code{apt-get(1)} calls with \protect\code{fakeroot(1)}.}
  \label{fig:debian-win-manual}
\end{figure}

\figurename~\ref{fig:debian-win-manual} shows analogous modifications of the Debian~10 build from \figurename~\ref{fig:debian-fail}. This needed more complex changes:
\begin{enumerate}

\item Disable \code{apt-get(1)}'s privilege dropping.

\item Update the package indexes. The base image contains none, so no packages can be installed without this update.

\item Install \code{pseudo}. (In our experience, the \code{fakeroot} package in Debian~10 was not able to install the packages we tested.)

\item Prepend \code{fakeroot(1)} to the install of \code{openssh-client}.

\end{enumerate}
Line 21 still has a warning. In our experience, Debian package management under \code{fakeroot(1)} occasionally complains about privileged operations failing, but these warnings do not stop the build.

These modifications leave us with working builds, but we want to build \emph{without} modifying any Dockerfiles.

\subsection{Auto-injection of \protect\code{fakeroot(1)}}

Unprivileged build of unmodified Dockerfiles can be done by automatically modifying the build. Charliecloud's \code{ch-image(1)} contains such an implementation. Its design principles are:
\begin{enumerate}

\item Be clear and explicit about what is happening.

\item Minimize changes to the build.

\item Modify the build only if the user requests it, but otherwise say what \emph{could} be modified.

\end{enumerate}
We outline Charliecloud's implementation in the context of the two running example Dockerfiles.

\subsubsection{Example: CentOS~7}

\begin{figure}
\begin{lstlisting}
$ ch-image build --force -t foo -f centos7.dockerfile
  1 FROM centos:7
will use --force: rhel7: CentOS/RHEL 7
  2 RUN ['/bin/sh', '-c', 'echo hello']
hello
  3 RUN ['/bin/sh', '-c', 'yum install -y openssh']
workarounds: init step 1: checking: $ command -v fakeroot > /dev/null
workarounds: init step 1: $ set -ex; if ! grep -Eq '\[epel\]' /etc/yum.conf /etc/yum.repos.d/*; then yum install -y epel-release; yum-config-manager --disable epel; fi; yum --enablerepo=epel install -y fakeroot;
+ grep -Eq '\[epel\]' [...]
+ yum install -y epel-release
[...]
Complete!
+ yum-config-manager --disable epel
[...]
+ yum --enablerepo=epel install -y fakeroot
[...]
Complete!
workarounds: RUN: new command: ['fakeroot', '/bin/sh', '-c', 'yum install -y openssh']
[...]
Complete!
--force: init OK & modified 1 RUN instructions
grown in 3 instructions: foo
\end{lstlisting}
  \listingspace
  \caption{Successful CentOS~7 build with unmodified Dockerfile. The \protect\code{fakeroot(1)} calls are almost identical to \figurename~\ref{fig:centos-win-manual} but automatically injected by \protect\code{ch-image}.}
  \label{fig:centos-win}
\end{figure}

We can build the CentOS~7 Dockerfile in \figurename~\ref{fig:centos-fail} with “\code{ch-image --force}” to enable the modifications; \figurename~\ref{fig:centos-win} shows the result. Like the other figures, this transcript shows the in-container commands that actually execute the \code{RUN} instructions, as well as new commands to implement the modifications. The basic steps are simple: (1)~test if the distribution is supported, (2)~install \code{fakeroot(1)} if needed, and (3)~augment \code{RUN} instructions that seem to need it. This Dockerfile works as follows.

First, the \code{FROM} instruction (lines~2–3) tells us that \code{ch-image} can use modifications requested with \code{--force}, though it doesn't yet know whether it \emph{will}. It tested known configurations and found a matching one called \code{rhel7}: the file \code{/etc/redhat-release} exists and its contents match the regular expression “\code{release 7\.}”. (This approach avoids executing a command within the container.)

Next, \code{ch-image} executes the first \code{RUN} instruction (lines~4–5) normally, because it doesn't seem to need modification. The second \code{RUN} instruction (line~6), contains the string “\code{yum}”, a keyword configured to trigger modification of the instruction.

Because this is the first modified \code{RUN}, \code{ch-image} must initialize \code{fakeroot(1)}. A configuration has a sequence of steps to do this (though for \code{rhel7} there is only one step), and each step contains two phases: a shell command to check if the step needs to be done, and a shell command to do the step. Here, the check (line~7) tests if \code{fakeroot(1)} is already installed in the image. In this example, \code{fakeroot(1)} is \emph{not} already present, so \code{ch-image} proceeds to the installation step (line~8), which for \code{rhel7} is a long pipeline with individual commands echoed for clarity (lines~9, 10, 13, 15). The pipeline installs EPEL if it's not already installed, but doesn't enable it because EPEL can cause unexpected upgrades of standard packages, and then it installs \code{fakeroot} from EPEL.  Files are \code{grep(1)}ed directly, rather than using \code{yum repolist}, because the latter has side effects, e.g.\ refreshing caches from the internet.

Finally, the instruction is modified by inserting “\code{fakeroot}” before the \code{/bin/sh} call that executes it. This would be repeated for later modifiable \code{RUN}s, if there were any.

If the user had \emph{not} specified \code{--force}, some of this would still occur. \code{ch-image} still looks for a matching configuration, and the keyword test is still done for \code{RUN} instructions. The result is that \code{ch-image} knows that that modifications are \emph{available}, which it suggests if the build fails. In fact, it suggested \code{--force} in a transcript line omitted from \figurename~\ref{fig:centos-fail}.

\subsubsection{Example: Debian~10}
\label{sec:debian-win}

\begin{figure}
\begin{lstlisting}
$ ch-image build --force -t foo -f debian10.dockerfile
  1 FROM debian:buster
will use --force: debderiv: Debian (9, 10) or Ubuntu (16, 18, 20)
  2 RUN ['/bin/sh', '-c', 'echo hello']
hello
  3 RUN ['/bin/sh', '-c', 'apt-get update']
workarounds: init step 1: checking: $ apt-config dump | fgrep -q 'APT::Sandbox::User "root"' || ! fgrep -q _apt /etc/passwd
workarounds: init step 1: $ echo 'APT::Sandbox::User "root";' > /etc/apt/apt.conf.d/no-sandbox
workarounds: init step 2: checking: $ command -v fakeroot > /dev/null
workarounds: init step 2: $ apt-get update && apt-get install -y pseudo
[...]
Setting up pseudo (1.9.0+git20180920-1) ...
[...]
workarounds: RUN: new command: ['fakeroot', '/bin/sh', '-c', 'apt-get update']
[...]
  4 RUN ['/bin/sh', '-c', 'apt-get install -y openssh-client']
workarounds: RUN: new command: ['fakeroot', '/bin/sh', '-c', 'apt-get install -y openssh-client']
[...]
--force: init OK & modified 2 RUN instructions
grown in 4 instructions: foo
\end{lstlisting}
  \listingspace
  \caption{Successful Debian~10 build with unmodified Dockerfile. The \protect\code{fakeroot(1)} calls are almost identical to \figurename~\ref{fig:debian-win-manual} but automatically injected by \protect\code{ch-image}.}
  \label{fig:debian-win}
\end{figure}

\figurename~\ref{fig:debian-win} shows a successful build of the Dockerfile in \figurename~\ref{fig:debian-fail} with \code{ch-image --force}; . In this case, \code{ch-image} selected configuration \code{debderiv} (line~3) because the file \code{/etc/os-release} contains the string “\code{buster}”. The first \code{RUN} is again left unmodified (lines~4–5).

The second \code{RUN} is deemed modifiable because it contains the string “\code{apt-get}”. Initialization for Buster takes two steps. First, \code{ch-image} tests whether the APT sandbox is disabled by configuration \emph{or} if user \code{_apt} is missing (line~7); it's not, so a configuration file (line~8) to disable it is added. Next, it tests for the availability of \code{fakeroot(1)} (line~9); it's not present, so check whether the package \code{pseudo} (line~10) is installed, which is in the standard repositories. It would be better not to update the package indexes with \code{apt-get update} because this is an operation with many side effects, but the base images ship with no indexes, so nothing can be installed until after this command.

The modified \code{RUN} is then executed (line~14); note that \code{ch-image} is not smart enough to notice that it's now redundant and could have been skipped. Finally, modify and execute the \code{RUN} which installs the problem package \code{openssh-client} (line~16–17).

\subsubsection{Production}

\code{ch-build --force} also works for various production applications at Los Alamos. For example, one application has integrated Charliecloud container build into its CI pipeline using a sequence of three Dockerfiles: the first installs and configures OpenMPI in a CentOS base image, the second installs the complex Spack~\cite{gamblin2015spack} environment needed by the application, and the third builds the application itself. Once built, the third image is pushed to a private container registry. The validation stage then pulls this image and uses it to run the application's test suite. Build and validate both run on supercomputer compute nodes using normal jobs, and the pipeline is coordinated by a separate GitLab server.

\section{Discussion}
\label{sec:discussion}

We have argued that HPC should minimize privilege for container image build, and we have detailed how container implementations can reduce privilege with user namespaces. We demonstrated two approaches with real products: mostly-unprivileged Type~\ref{ct:priv-user} rootless Podman and fully-unprivileged Type~\ref{ct:unpriv-user} Charliecloud. We now compare the two approaches, outline future work, and discuss how these methods for container builds can impact the field of HPC more broadly.

\subsection{Mostly vs.\ fully unprivileged build}

To demonstrate Type~\ref{ct:priv-user} container build via privileged user namespaces, we used rootless Podman. The main advantage of this approach is that it retains file ownership in the container image and presents no UID/GID oddities to distribution tools, allowing traditional builds to work for otherwise-unprivileged users. The main disadvantage is reliance on correct implementation and configuration of the privileged helper tools.
One current problem with rootless Podman is that its user namespaces implementation clashes with shared filesystems, including default-configured Lustre, GPFS, and NFS. The issue is not a container problem per se, but rather the user extended attributes (xattrs) Podman uses for its ID mappings.

To demonstrate Type~\ref{ct:unpriv-user} container build via unprivileged user namespaces, we used Charliecloud. The main advantage here is that the entire build process is fully unprivileged; all security boundaries remain within the Linux kernel. However, the Charliecloud build process is more complex; current disadvantages include:
\begin{enumerate}

  \item The \code{fakeroot(1)} wrapper is required for a large fraction of builds, which introduces another layer of indirection. While this can be automated, the wrapper must be applied separately to each \code{RUN} instruction, and it's extremely difficult to tell precisely which \code{RUN}s really need it.

  \item The resulting image is slightly different. All files must be owned by a single user (the unprivileged user who did the build), and the image cannot contain privileged special files such as devices. On push, Charliecloud changes ownership for all image files to root:root and clears setuid/setgid bits, to avoid leaking site IDs. Charliecloud currently adds two further complications: \code{fakeroot(1)} is installed into the image, and images are single-layer, in contrast to other implementations that push images as multiple layers. In general, however, these differences should not impact the behavior of HPC applications.

  \item Charliecloud lacks a per-instruction build cache, in contrast to other leading Dockerfile interpreters including Podman and Docker. This caching can greatly accelerate repetitive builds, such as during iterative development.

\end{enumerate}

On balance, neither solution currently fits all HPC use cases. Generally, if preserving file ownership within container images is more important than the security cost of privileged helper tools, Type~\ref{ct:priv-user} builders like Podman are likely a good choice; if the opposite, then Type~\ref{ct:unpriv-user} builders like Charliecloud are likely preferred.

\subsection{Future work}

Both Type~\ref{ct:priv-user} and Type~\ref{ct:unpriv-user} container build have high-quality, actively developed implementations available today. However, there are still several opportunities for improvement.

\subsubsection{Recommendations for Type~\ref{ct:priv-user} implementations}


First, adding a Type~\ref{ct:unpriv-user} mode to Type~\ref{ct:priv-user} implementations that already use user namespaces is straightforward; doing so would make an implementation more flexible and adaptable. Second, shared filesystems are critical for HPC, so the xattrs support is still necessary. For Lustre, support needs to be enabled on both the metadata server and storage targets. For NFS, changes in Linux~5.9 along with NFSv4~\cite{naik2017xattrs} server bring in xattrs support, though at the time of writing these changes have yet to be extended into distributions common in HPC. We have not evaluated GPFS support at the time of writing. To our knowledge, none of these configurations have yet been tested with Type~\ref{ct:priv-user} containers.

\subsubsection{Recommendations for Type~\ref{ct:unpriv-user} implementations}

In addition to Charliecloud-specific improvements in progress like layers and build caching, we identify three more general refinements here:

\begin{enumerate}

  \item Fix \code{fakeroot(1)}. Not all implementations can install all packages; characterize the scope of the problem and address it. With a sufficiently robust \code{fakeroot(1)}, Type~\ref{ct:unpriv-user} could become a drop-in replacement for Type~\ref{ct:priv-nouser}, eliminating the need for the Type~\ref{ct:priv-user} privilege compromise.

  \item Preserve file ownership. \code{fakeroot(1)} does track this so it can tell consistent lies; the information could be extracted and used when exporting or pushing images. For example, Charliecloud uses the Python \code{tarfile} package and could create layer archives that reflect \code{fakeroot(1)}'s database rather than the filesystem.

  \item Move \code{fakeroot(1)}. Rather than installing in the image itself, the wrapper could be moved into the container implementation. This would simplify it and also ease the prior item. At least one \code{fakeroot(1)} implementation already has a \code{libfakeroot}~\cite{man2019fakeroot}.

\end{enumerate}

\subsubsection{Recommendations for distributions}
\label{sec:rec-distros}

Beyond improvements to container build implementations, package managers and other distribution tools could improve support for unprivileged users, eliminating the need for Type~\ref{ct:priv-user} privileged code or Type~\ref{ct:unpriv-user} wrappers. Given increasing use of containers across the computing industry, there may be compelling reasons to support them even beyond HPC. The complexity of this effort should be better understood, and edge cases and older distributions mean container build tools will need their own low-privilege build support for some time.

\subsubsection{Recommendations for ID maps}

Consider that \code{fakeroot(1)} manages user identity and file ownership in user space; in contrast, the user namespace ID maps rely on the kernel to track this state. Currently, Type~\ref{ct:priv-user} kernel ID maps require helper tools like \code{newuidmap(1)} and \code{newgidmap(1)}; they are privileged because this is a security-sensitive operation and to ensure system configuration (\code{/etc/subuid} and \code{/etc/subgid}) is respected. Future kernel versions could provide other mechanisms to expand the utility of unprivileged maps. For example, supplemental groups could become mappable, general policies could be implemented such as “host UID maps to container root and guaranteed-unique host UIDs map to all other container UIDs”, or the kernel could manage the fake ID database with actual files stored as the invoking user.

\subsubsection{Recommendations for standards}

Above, we argued that file ownership for HPC application containers is usually not needed, and instead it's mostly an artifact of legacy distribution tools used for building images. A flattened file tree where all users have equivalent access, like that produced by Charliecloud or Singularity's SIF image format, is sufficient and in fact advantageous for most HPC applications. We suspect many non-HPC container use cases have similar requirements. A potential extension to the OCI specification and/or the Dockerfile language is explicit marking of images to disallow, allow, or require them to be ownership-flattened.

\subsection{Impact on HPC}

This paper argues that low-privilege container build will have significant impact for the science enabled by HPC, because these capabilities allow containers to be built directly on the HPC resources where they will be run. This property is especially important for HPC developers and users due to the specialized nature of supercomputing hardware, which makes building on dissimilar laptops, workstations, or CI/CD virtual machines undesirable.

We expect these capabilities to also improve DevOps modernization of scientific computing application development and deployment. Specifically, DevOps coupled with low-privilege container build will allow CI/CD pipelines to execute directly on supercomputing resources such as login/front-end and compute nodes, perhaps in parallel across multiple supercomputers or node types to automatically produce specialized container images. For future exascale supercomputers in particular, this user-based build capability, coupled with automated CI runners will enhance development, testing, and deployment of HPC applications directly on the appropriate resources. This integration should reduce not only the effort of developing or porting complex codes on new systems, but also that of managing development resources matching the architecture of one site's supercomputer across many other sites.

%



\begin{acks}

  Eduardo Arango, Dan Walsh, Scott McCarty, and Valentin Rothberg of Red Hat, Inc.\ helped us understand rootless Podman. Peter Wienemann of the Debian Project clarified APT privilege dropping. Michael Kerrisk of man7.org Training and Consulting provided corrections for the namespaces discussion. (All remaining errors are of course ours alone.) Kevin Pedretti and the ATSE team helped implement Astra's container workflow.

  This work was supported in part by the Exascale Computing Project (17-SC-20-SC), a collaborative effort of the U.S.\ Department of Energy (DOE) Office of Science and the National Nuclear Security Administration (NNSA); the Advanced Simulation and Computing Program (ASC); and the LANL Institutional Computing Program, which is supported by the U.S.\ DOE’s NNSA under contract 89233218CNA000001. Sandia National Laboratories is a multimission laboratory managed and operated by National Technology and Engineering Solutions of Sandia, LLC., a wholly owned subsidiary of Honeywell International, Inc., for the U.S.\ DOE’s NNSA under contract DE-NA0003525. LA-UR~21-23314; SAND2021-4332~O.

\end{acks}


\bibliographystyle{ACM-Reference-Format}
\bibliography{refs-cooked.bib}

\end{document}